\documentclass[preprint,review,12pt,sort&compress]{elsarticle}
\usepackage{graphicx}% Include figure files
\usepackage{dcolumn}% Align table columns on decimal point
\usepackage{bm}% bold math
\usepackage{multirow}
\usepackage{hyperref}% add hypertext capabilities
\usepackage{booktabs}

\hypersetup{
colorlinks=true,% false: boxed links; true: colored links
linkcolor=blue,% color of internal links
citecolor=blue,% color of links to bibliography
urlcolor=blue% color of external links
}

\usepackage{chemformula}
\usepackage[output-decimal-marker={.}]{siunitx}

\DeclareSIUnit\angstrom{\text{\AA}}

\makeatletter
\def\ps@pprintTitle{%
  \let\@oddhead\@empty
  \let\@evenhead\@empty
  \def\@oddfoot{\reset@font\hfil\thepage\hfil}
  \let\@evenfoot\@oddfoot
}
\makeatother

\begin{document}

\begin{frontmatter}



\title{Combined \textit{ab initio} and experimental study of phosphorus-based\\
anti-wear additives interacting with iron and iron oxide} %% Article title


\author[inst1]{Francesca Benini}

\affiliation[inst1]{organization={Dipartimento di Fisica e Astronomia, Università di Bologna},%Department and Organization
            addressline={Viale Berti Pichat 6/2}, 
            city={Bologna},
            postcode={40127}, 
            %state={State One},
            country={Italy}}

\author[inst1]{Paolo Restuccia}

\author[inst2]{Sophie Loehlé}

\affiliation[inst2]{organization={TotalEnergies, OneTech Fuels\&Lubricants, Research Center Solaize},%Department and Organization
            addressline={Chemin du Canal BP 22}, 
            city={Solaize},
            postcode={69360}, 
            %state={State One},
            country={France}}

\author[inst2]{Quentin Arnoux}

\author[inst1]{M. Clelia Righi\corref{mycorrespondingauthor}}
\ead{clelia.righi@unibo.it}
\cortext[mycorrespondingauthor]{Corresponding author}

%% Abstract
\begin{abstract}
The performance of phosphorus-based lubricant additives is governed by their adsorption, stability, and reactivity at the metal interface. In this study, we investigate the adsorption behavior and tribochemical stability of three additives: Octyl Acid Phosphate (OAP), Dibutyl Hydrogen Phosphite (DBHP), and Amine Neutralized Acid Phosphate (ANAP). These additives are studied on iron and hematite surfaces using both \textit{ab initio} calculations and experimental analyses on steel. 
Simulations revealed that ANAP exhibited the strongest adsorption on iron, followed by DBHP, while OAP showed weaker interactions, though its chemisorption was enhanced on hematite via hydrogen loss. Under tribological conditions, the DBHP phosphite dissociated more readily than the other two phosphates molecules due to its lower phosphorus coordination, as confirmed by bond order analysis. Quartz crystal microbalance (QCM) measurements indicated significant differences in adsorption behavior across temperatures, with DBHP forming stable deposits, while ANAP exhibited poor retention, in agreement with \textit{ab initio} molecular dynamics simulations. X-ray photoelectron spectroscopy (XPS) confirmed DBHP’s strong chemisorption and molecular dissociation, leading to increased phosphorus deposition. OAP, despite forming a phosphorus-based layer, caused a reduction in Fe oxide, consistent with its hydrogen release mechanism observed in simulations. These findings highlight the critical role of molecular structure and oxidation state in tribofilm formation and stability. Understanding these interactions at the atomic level provides valuable insights for designing high-performance lubricant additives for extreme operating conditions.
\end{abstract}

\begin{graphicalabstract}
\includegraphics[width=\columnwidth]{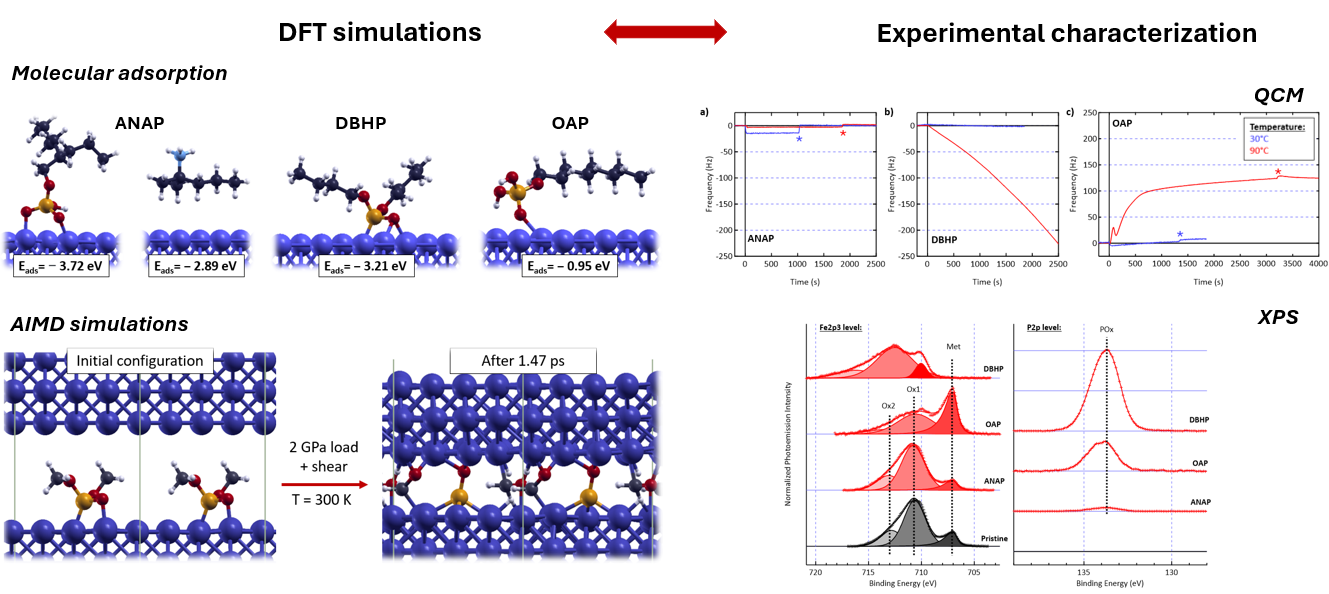}
\end{graphicalabstract}

%\item XXX
%\item XXX
%\item XXX
%\item XXX

%% Keywords

\end{frontmatter}



\section{Introduction}
\label{sec:intro}

Reducing friction and wear in automotive applications is critical for boosting engine and transmission efficiency. In this context, liquid lubricants are essential and they are carefully engineered to modify friction and protect the surfaces of moving components against damage under extreme pressure and temperature. Their design can significantly improve the energy efficiency in vehicles, while also adhering to strict environmental regulations and taking advantage of cutting-edge technological advancements~\cite{Spikes2015}.

Modern liquid lubricants are formulated by blending base oils with a range of molecular additives, each tailored to meet specific performance standards for different vehicle components. 
Phosphorus-based additives, in particular, play a key role due to their unique ability to form protective layers on metal surfaces, reducing friction and wear under high-stress conditions~\cite{spikes2008low}. Under conditions of heat and pressure, these additives undergo chemical reactions that produce thin, durable layers known as tribofilms, which bond tightly to metal surfaces. These tribofilms help reduce metal-to-metal contact, thus minimizing wear and prolonging the lifetime of materials.

In this context, phosphorus-based molecules stand out as some of the most commonly used additives in liquid lubricants. 
Among these, phosphates and phosphites, characterized by phosphorus oxidation states of
$+5$ and $+3$, respectively, serve distinct roles: organic phosphites primarily function as friction modifiers, whereas phosphates are more effective as anti-wear (AW) additives.~\cite{riga2001organophosphorus,gao2005reaction,gao2004reaction,loehle2018ab}. Within this class of chemically active compounds, extreme-pressure (EP) additives are essential in boundary lubrication, where the oil thickness becomes critically thin, risking direct metal-to-metal contact~\cite{spikes2008low,tang2014review,furlong2007understanding}. These protective qualities and durability make phosphorus-based additives essential for maintaining the efficiency and longevity of critical engine and transmission components.

In the boundary lubricating regime, scuffing can emerge by forming local welds between sliding surfaces, causing surges in asperity friction and resulting in damage. It frequently occurs in heavily loaded, high-speed sliding contacts such as gears, sliding cam-followers, and roller bearings~\cite{dyson1975scuffing}. Many factors can trigger this phenomenon. For instance, the breakdown of the elastohydrodynamic (EHL) film and tribofilms that separate rubbing surfaces is a common cause. Other changes in contact conditions like increasing load, speed, temperature, or the introduction of solid contaminants can lead to scuffing. Nowadays, the resistance to scuffing is particularly relevant in the development of lubricants for high-speed electric vehicle transmissions~\cite{li2022determination}.

Extensive research has explored the factors contributing to EHL film breakdown. However, the processes governing tribofilm formation, removal, and their impact on scuffing remain only partially understood~\cite{bowman1996review}. Tribofilms generated by EP and AW additives, especially those based on phosphorus, have been shown to play a crucial role in mitigating scuffing. Their composition and durability are critical in controlling scuffing, as a stable tribofilm can resist breakdown even under extreme operating conditions~\cite{ueda2022situ}. Understanding the dynamics of tribofilm formation and removal is essential for advancing scuffing resistance strategies, paving the way for creating more resilient lubricants in automotive applications.

This study aims to offer a comprehensive insight into the molecular-level interaction between phosphorus-based lubricant additives and iron, specifically emphasizing their role in preventing scuffing. Combining experimental data obtained by Quartz Crystal Microbalance (QCM) and X-ray photoelectron spectroscopy (XPS) techniques with \textit{ab initio} calculations can help uncover the tribochemical processes leading to scuffing and ways to prevent it.
In particular, we compared three different molecules typically employed as anti-wear additives in commercial formulations: Dibutyl Hydrogen Phosphite (DBHP), Octyl Acid Phosphate (OAP) and an amine-neutralized acid phosphate (ANAP) (Fig.~\ref{fig:mols}). Together with their interaction with Fe and hematite, we investigated the stability of their internal bonds to predict their effect in tribological applications.

\section{Methods}
\label{sec:methods}

\subsection{Computational methods}
The calculations were performed utilizing spin-polarized Density Functional Theory (DFT) implemented in version 7.2 of the Quantum Espresso suite~\cite{Giannozzi2009, Giannozzi2017, Giannozzi2020}. The exchange-correlation functional was specified using the generalized gradient approximation (GGA) within the Perdew-Burke-Ernzerhof (PBE) parametrization~\cite{Perdew1996}. The use of this functionals class is common in computational tribology and has been compared with other setups, resulting to be efficient in producing results confirmed by experiments, especially when combined with Van der Waals corrections~\cite{losi2020superlubricity, benini2024zinc}. 
In this work, two distinct computational setups were employed, one for the (001) hematite surface description and another for the remaining calculations.

For the (001) hematite surface description, we employed DFT+U~\cite{anisimov1991band} with a Hubbard parameter set to 4.2 eV, a choice aligned with previous studies on adsorption on hematite~\cite{nguyen2013water, wang2018initial} and specifically suitable for the Vanderbilt ultrasoft pseudopotentials used. The kinetic energy cutoff for the wave functions was established at 50 Ry, with a charge density cutoff of 400 Ry. The spin of \textit{d} electrons localized on Fe atoms was assigned to depict the antiferromagnetic nature of this material~\cite{dzade2014density}. Additionally, a Gaussian smearing with a width of 0.02 Ry was applied.

For all other calculations, we employed standard DFT. The kinetic energy cutoff for the wave functions was set to 40 Ry, with a charge density cutoff of 320 Ry. A Gaussian smearing with a width of 0.001 Ry was incorporated to enhance the description of occupations around the Fermi level. The electronic configuration of atoms was represented by ultrasoft pseudopotentials within the RRKJ parametrization~\cite{rkkj}. We computed the adsorption energies by considering calculations with and without the addition of van der Waals corrections. In the former case, we exploited a customized modification of the Grimme D2 dispersion correction scheme, denoted as D$_{NG}$~\cite{marquis2022nanoscale, grimme2006semiempirical}. This modification deviates from the standard D2 primarily in terms of the C$_6$ coefficient and the van der Waals radius $R_0$ of the Fe atoms. Notably, these parameters are substituted with those of Ar (argon), the noble gas preceding Fe.

To investigate the molecular adsorption on iron surfaces, we utilized large $6 \times 8$ orthorhombic supercells to prevent unintended lateral interactions among replicas. Each slab represented a (110) Fe surface consisting of four layers containing 48 atoms each. Additionally, we incorporated a substantial void in each cell to ensure a vertical separation of at least 10 \r{A} between replicas. The resulting supercell dimensions were 17.10 \r{A} $\times$ 16.12 \r{A} $\times$ 30.78 \r{A}, determined after optimizing a smaller $2 \times 2$ cell. Given the large size of the supercell, we optimized it by sampling solely at the $\Gamma$ point. We identified the most favorable binding site and molecular orientation using a software to automatize the adsorption studies, Xsorb~\cite{Xsorb}, and computed the adsorption energies using the formula:

\begin{equation}
    E_{ads} = E_{tot} - (E_{sub} + E_{adsorbate})
    \label{eads}
\end{equation}

where $E_{tot}$ is the energy of the optimized system comprising the substrate and the adsorbed molecule/atom, and $E_{sub}$ ($E_{adsorbate}$) is the energy of the isolated substrate (adsorbate).

We also performed \textit{ab initio} Molecular Dynamics Simulations (AIMD) using a modified version of the Quantum Espresso code to simulate tribological conditions~\cite{Ramirez2020} with the same computational parameters as for the static calculations. During the AIMD simulations, the bottom atomic layer of the lower slab was fixed, while a constant velocity of 200 m/s was imposed to the uppermost layer of the counter-surface. We also applied a vertical force to the atoms of the uppermost layer to impose an external normal pressure of 2 GPa. In the initial configurations, a single molecule of any additive type considered in this study was adsorbed on the iron substrate in its most stable configuration. The system underwent relaxation under an applied pressure of 500 MPa, followed by a equilibration at a constant temperature of 300 K with a timestep of 1 fs. This setup is analogous to the one employed to study trimethylphosphite (TMP) interacting with iron~\cite{loehle2018ab}.

In order to compare the strength of different bonds within the molecule, we evaluated the energetic costs associated to different cuts. To do so, we performed static calculations at frozen ionic positions of the isolated fragments in vacuum. The fragmentation energy $\Delta E$, corresponding to the energy required to break each bond, is calculated as:

\begin{equation}
    \Delta E = \sum_{n=1}^N E_{f_n} - E_{mol}.
    \label{deltae}
\end{equation}

 in which $N$ denotes the number of fragments obtained for each of the investigated cuts, $E_{f_n}$ is the calculated energy of each fragment (identified by $n$), and $E_{mol}$ is the energy of the isolated molecule.
Finally, we performed bond order analysis and crystal orbital overlap population exploiting the software LOBSTER~\cite{lobs,lobs2,lobs3,lobster}. To do so, we evaluated the integrated Crystal Orbital Bond Index (ICOBI) and the integrated Crystal Orbital Overlap Population (ICOOP).
To account for the charged nature of the two components of the ANAP, we set an initial $-$1 ($+$1) charge on the O (H) atoms in component a (b).

\subsection{Experimental methods}
To properly characterize the film formed on the metallic surfaces, Quartz Crystal Microbalance (QCM) has been used to deposit the AW additives of the lubricant. X-ray Photoelectron Spectroscopy (XPS) is then performed to determine film composition and chemistry.
\subsubsection{Quartz Crystal Microbalance sample preparation}
For this study, we used quartz crystal sensors coated with stainless steel SS2343 (QSX304, BiolinScientific AB, Sweden). The quartz crystal sensors are mounted into the QCM-D flow modules (QSense Explorer, BiolinScientific, Sweden). Two types of chambers are used, one for ambient temperatures (30 °C) and the other for high temperatures (90 °C). The oscillation frequency ($f$) of the quartz crystals was measured for the fundamental frequency (5 MHz) and for six overtones in air until the base line was stabilized ($\Delta f$ over 10 min less than 2.0 Hz). Then, base oil was pumped into the flow module at a flow rate of 0.2 ml/min. After frequency stabilization, a solution of AW (0.1$\%$w DBHP, 0.5$\%$w OAP and 1$\%$w ANAP) in base oil is pumped for approximately 30 min for the evaluation of their interaction with the stainless-steel surface, followed by pumping pure base oil for rinsing until the signal stabilized. In this apparatus, the sensor is a thin quartz disc with gold electrodes evaporated on each one of its opposing surfaces. It oscillates at its resonance frequency when the corresponding alternating current is applied. Molecular adsorption at the electrode surface causes a change in the sensor mass that results in a linear change of its oscillation frequency when the adsorbate is solid-like. The relation is shown in Eq.\ref{eq1}, where $m$ is the Sauerbrey sensed mass (in mg m$^{-2}$) calculated from the third overtone, $n$ being the overtone number (dimensionless), the measured frequency $f$ (in Hz) and the constant $C$ (in this case 0.177 mg m$^{-2} Hz^{-1}$)\cite{huang2016surface,wang2018surface,rollmann2004first}. Thus, the real-time frequency changes measured by the QCM setup allow the appreciation of the ongoing surface phenomena.
\begin{equation}
    \Delta m = -\frac{1}{n}C\Delta f
\label{eq1}
\end{equation}
For data visualization, a “Savitzky-Golay” filter has been used to smooth the data.

\subsubsection{X-Ray Photoelectron Spectroscopy}
XPS analysis was performed with a Thermo NEXSA spectrometer (Thermo Electron, USA) to obtain the surface composition through survey and high-resolution spectra of the C1s, O1s, P2p, N1s, Fe2p core levels. The X-ray excitation was the Al-K$\alpha$ line at 1486.7 eV. The detection of the photoelectrons was perpendicular to the sample surface. The spot size was 400 $\mu$m. Flood gun was turned on for charge compensation in all adsorbate deposit samples. Material homogeneity has been checked by  studying 2 different spots on each sample.

\section{Results}
\begin{figure}[htpb]
\includegraphics[width=\columnwidth]{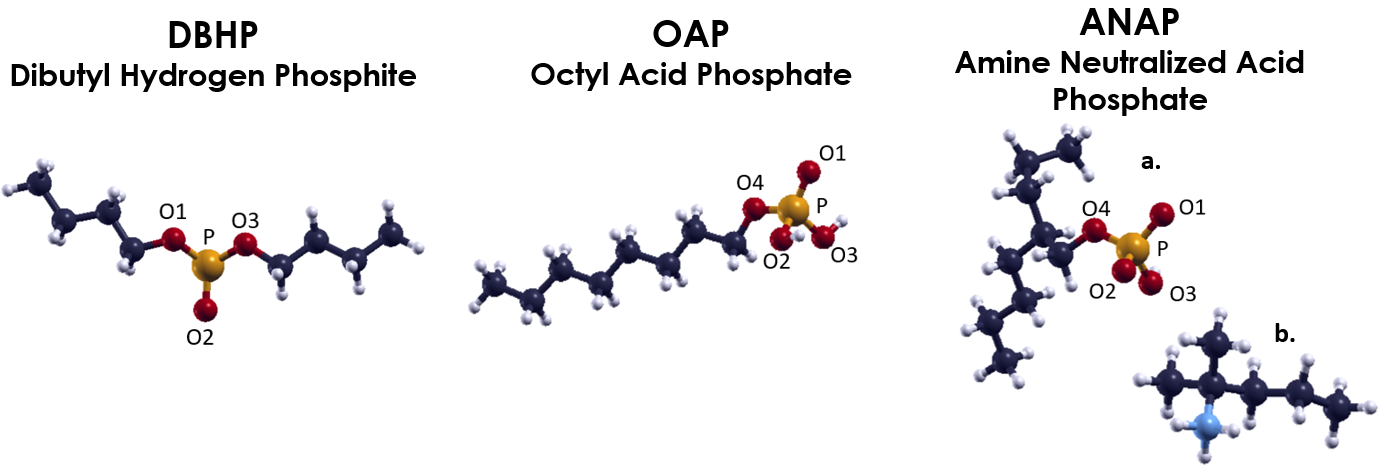}
\caption{\label{fig:mols} The three molecules studied in this work. From now on, white atoms represent H, black C, yellow P, red O and blue N, respectively.}
\end{figure}

The three molecules considered in this study, namely dibutyl hydrogen phosphite (DBHP), octyl acid phosphate (OAP) and amine neutralized acid phosphate (ANAP), are shown in Fig.~\ref{fig:mols}. 

DBHP is characterized by a phosphite unit with two terminating primary butyl chains with formula C$_8$H$_{18}$O$_3$P. OAP is a monoalkyl phosphoric acid ester with formula C$_8$H$_{19}$O$_4$P whereas the ANAP is an ionic liquid composed of a phosphate anion with formula C$_9$H$_{20}$O$_4$P (a.) and an amine cation C$_6$H$_{16}$N (b.).

\subsection{\label{sec:stat}Adsorption and dissociation on Fe}

\subsubsection{Adsorption configurations and energies on iron}
\begin{figure}[htpb]
\includegraphics[width=\columnwidth]{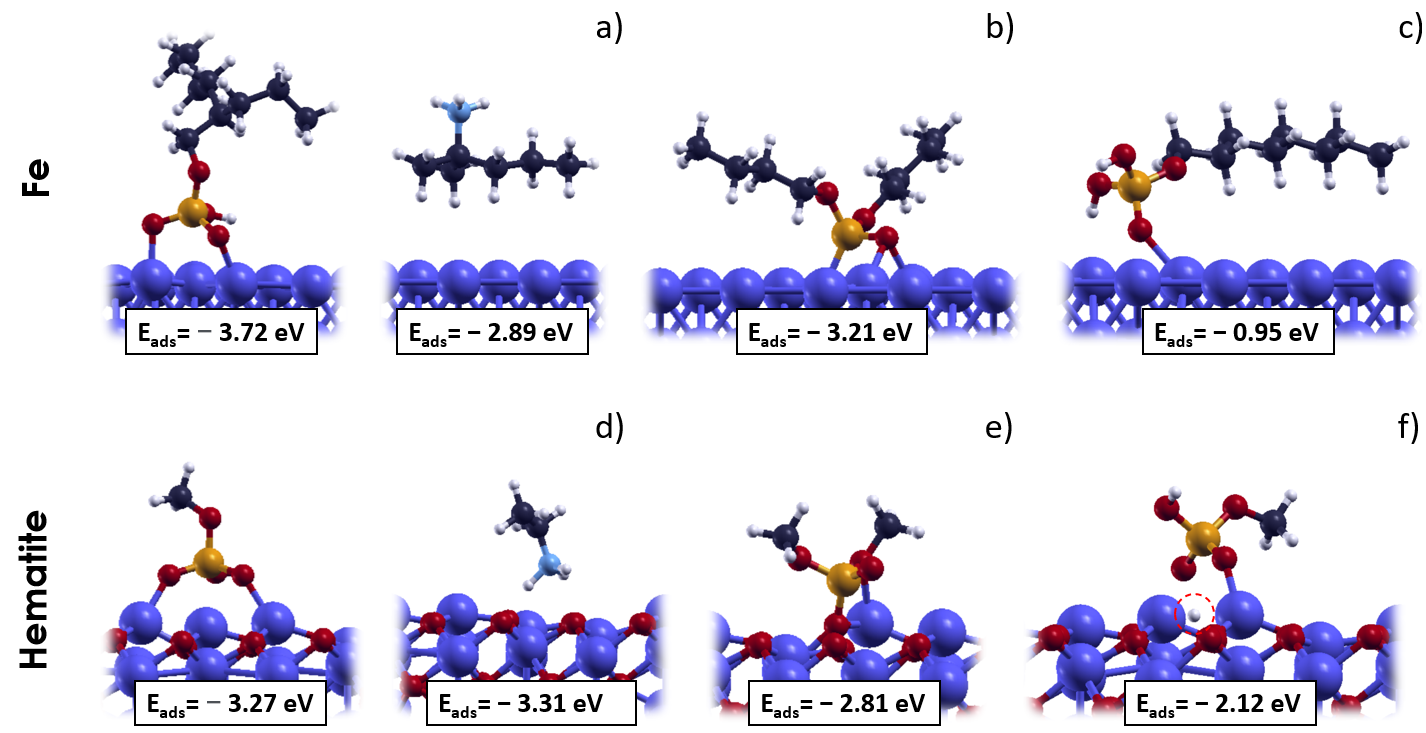}
\caption{\label{fig:ads} Adsorption of ANAP (a), DBHP (b), and OAP (c) on Fe (upper panel) and of their simplified versions, denoted with *, on hematite (lower panel). The adsorption energy values are reported in the picture.}
\end{figure}

We initially simulated the adsorption of the molecules on bare iron. Once we identified the best adsorption configuration using Xsorb, we computed the associated adsorption energy $E_{ads}$ and found that the strongest interaction is obtained for the amine neutralized acid phosphate, followed by DBHP. OAP interacted more weakly with iron. The optimal adsorption configurations are reported in panels a) to c) of Fig.~\ref{fig:ads}. 
We also performed all the adsorption calculations accounting for vdW interactions, with the results reported in Tab.~\ref{tab:en}. The dispersion interactions lead to an energy decrease for all configurations, while maintaining the energy trends and the geometrical configurations. 

\begin{table}[htpb]
\begin{center}
\caption{\label{tab:en} Adsorption energies (in eV) of the considered molecules on iron calculated without and with van der Waals corrections, respectively.}
\begin{tabular}{ c c c }
\hline
\hline
 & E$_{ads}$ (no vdW) & E$_{ads}$ (vdW) \\
\hline
ANAP a.     & $-$3.72                 & $-$4.37     \\
DBHP        & $-$3.21                 & $-$3.67     \\
ANAP b.     & $-$2.89                 & $-$3.24     \\
OAP         & $-$0.95                 & $-$1.62     \\
\hline                       
\end{tabular}
\end{center}
\end{table}

\begin{figure}[htpb]
\includegraphics[width=0.8 \columnwidth]{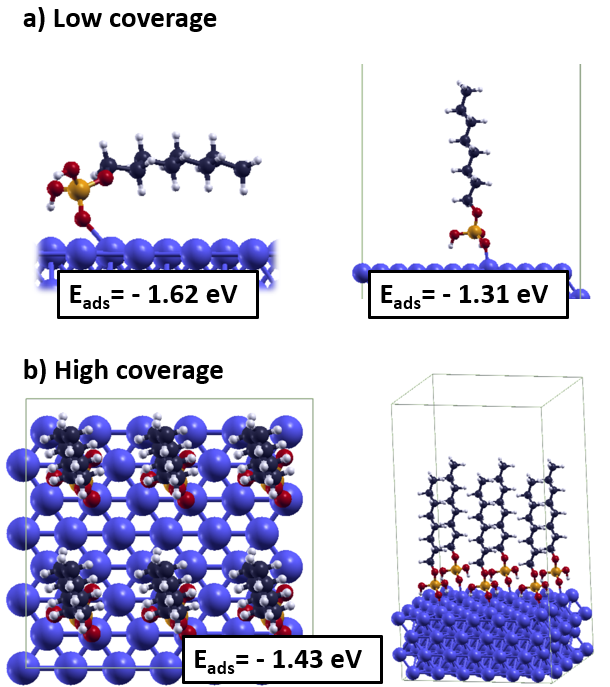}
\caption{\label{fig:mono} Comparison of the adsorption energy per molecule for OAP at low (panel a) and high (panel b) coverages. The energy values take into account the VdW corrections.}
\end{figure}

\subsubsection{Effects of substrate oxidation}
We then simulated the adsorption of the additives on hematite in order to find out how much the substrate oxidation can affect the binding strength of these molecules. Due to the high computational cost associated with hematite simulation, we reduced the size of the molecules by replacing the long alkyl chains with methyl groups, allowing to reduce the lateral size of the substrate supercell. This approach has been already adopted by our group to study the adsorption and dissociation of lubricant additives (i.e., MoDTC, ZDDP) over different substrates~\cite{peeters2019characterization, PEETERS2022153947}. Those studies showed that the energetics for molecules having methyl groups and longer alkyl chains in the ligand units are similar as the chemisorption strength is governed by the functional group. The shortened versions of the molecules are here denoted with a star symbol (*).

\begin{table}[htpb]
\begin{center}
\caption{\label{tab:ener} Adsorption energies (in eV) of the molecules with shorter chains on Fe(110) and hematite, respectively.}
\begin{tabular}{ c c c }
\hline
\hline
 & E$_{ads}$ (Fe) & E$_{ads}$ (Fe$_2$O$_3$) \\ \hline
*ANAP a.        & $-$3.91                       & $-$3.27                    \\
*DBHP           & $-$3.16                       & $-$2.81                    \\
*ANAP b.        & $-$2.70                       & $-$3.31                    \\
*OAP            & $-$0.83                       & $-$2.12                    \\
\hline         
\end{tabular}
\end{center}
\end{table}

As can be seen in the first column of Tab.~\ref{tab:ener}, the adsorption energies obtained for the shortened molecules on Fe(110) surface are in agreement with those obtained for the longer-chain molecules (indicated in the upper panel of Fig.~\ref{fig:ads}), confirming the minor role played by alkyl chains compared to the functional groups of the molecules~\cite{peeters2019characterization,benini2024zinc}.
The adsorption configurations on hematite are reported in the lower panel of Fig.~\ref{fig:ads}, including the adsorption energies. These values are compared with those obtained for the same shortened molecules on iron in Tab.~\ref{tab:ener}. The overall adsorption trend between the two substrates is maintained, with the largest difference arising in the case of *OAP. This molecule, which was found to weakly interact with the Fe substrate, now appears to lose an H atom which is chemisorbed to the hematite substrate, enhancing the interaction and leading to $E_{ads}$ = $-$2.12 eV. Indeed, Fe oxides are known to interact strongly with molecules which are capable of losing H atoms, inducing the removal of O atoms from the surface to form H$_2$O~\cite{spreitzer2019reduction}. This will be further explained in this context by the XPS results combined with the QCM experiments in the last section. \\

%%%VALUTARE SE METTERE I PROSSIMI DUE PARAGRAFI IN SUPPLEMENTARY
\subsubsection{Effects of molecular coverage}
For specific molecules characterized by a reactive head (``head group'') and a long alkyl chain (``tail group''), the energy gain associated to the formation of a self-assembled monolayer can lower the adsorption energy on a substrate. Experimental results suggest that this may be the case for the OAP, which is able to form a self-assembled monolayer at high concentrations on cerium conversion coating on galvanized steel~\cite{Kobayashi_2006}.
For this reason, even though a vertical orientation is energetically less favorable for a single OAP molecule adsorbed on Fe compared to the horizontal one ($-$0.86 eV vs $-$0.95 eV), we tested the adsorption energy per molecule when the coverage is increased for both orientations. 

Including vdW corrections for a better description of the intermolecular interaction, we obtained $E_{ads}$ = $-$1.31 eV for a single vertical molecule adsorbed on the surface and $E_{ads}$ = $-$1.43 eV per molecule when 6 molecules were simultaneously adsorbed on the same supercell (Fig.~\ref{fig:mono}). These findings suggest that, at higher coverages approaching a self-assembled monolayer, the vertical orientation of OAP becomes energetically comparable to the horizontal configuration ($E_{ads} = -1.62$ eV).

\subsection{\label{sec:dyn}Molecular dynamics simulations} 
\begin{figure}[htpb]
\includegraphics[width=\columnwidth]{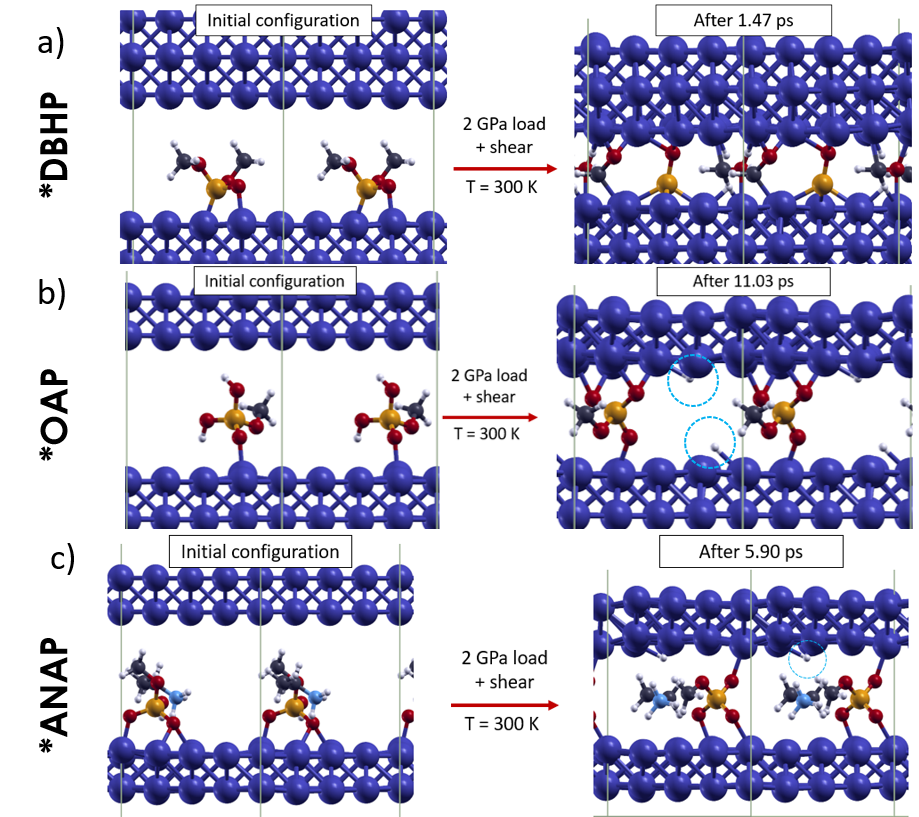}
\caption{\label{fig:dyn} AIMDs results for the three systems investigated: *DBHP (panel a), *OAP (b), and *ANAP (c). On the left (right), the initial (reported simulation time) configurations are shown.}
\end{figure}

We performed AIMD simulations to evaluate the most likely dissociation paths of OAP, DBHP, and ANAP adsorbed on a metallic substrate. To accelerate and enhance the chemical reaction processes, we decided to simulate each molecule confined within an iron interface in tribological environments (i.e., under applied temperatures, normal loads, and shear stresses), which are the typical operating conditions for these compounds. Key snapshots of the simulations are shown in Fig.~\ref{fig:dyn}. The total simulation time is equal to 11 ps. To reduce computational costs, we employed the shortened version of the molecules.

It is immediate to see in Fig.~\ref{fig:dyn}a that *DBHP had the weakest bonds to break, thus exhibiting the highest reactivity in tribological conditions. In particular, its behaviour was similar to trimethyl phosphite, thanks to their similar chemical structures, as observed previously by our group~\cite{loehle2018ab}. Specifically, the alkoxy chains were detached from the head group and chemisorbed onto the substrate due to the mechanical stresses applied in the system, leaving the central \hbox{P--O} fragment interacting with the iron slabs. Conversely, bonds within *OAP demonstrated a higher stability under these harsh conditions and did not dissociate (Fig.~\ref{fig:dyn}b). Nevertheless, the hydrogen atoms in the OH group dissociated and adsorbed onto the surface, in analogy to what we found in the static calculations performed on hematite. Finally, also the P in the *ANAP head group did not dissociate (Fig.~\ref{fig:dyn}c), likely due to the synergistic effect of its two ions. Like *OAP, *ANAP only loses hydrogen atoms that chemisorb onto the surface.

The qualitative analysis of the dissociation paths suggested that a higher coordination number for the central phosphorus with oxygen atoms (as in OAP and the phosphate anion in ANAP) helps in stabilizing the phosphorus-based molecules compared to those coordinated with only three oxygen atoms (as in DBHP and trimethyl phosphite).

\subsection{\label{sec:bond}Strength of  molecular bonds}

\begin{figure}[htpb]
\includegraphics[width=\columnwidth]{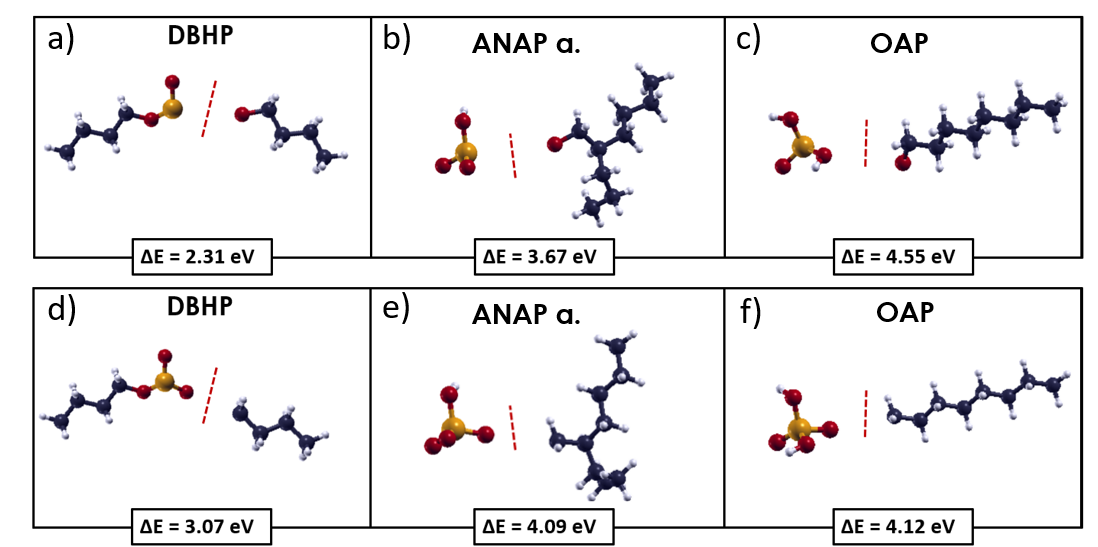}
\caption{\label{fig:chain} In the upper panel (a to c), alkoxy chain removal energy costs for DBHP (a), amine neutralized acid phosphate (b) and OAP (c). In the lower panel (d to f), alkyl chain removal energy costs for the same molecules.}
\end{figure}

\begin{figure}[htpb]
\includegraphics[width=\columnwidth]{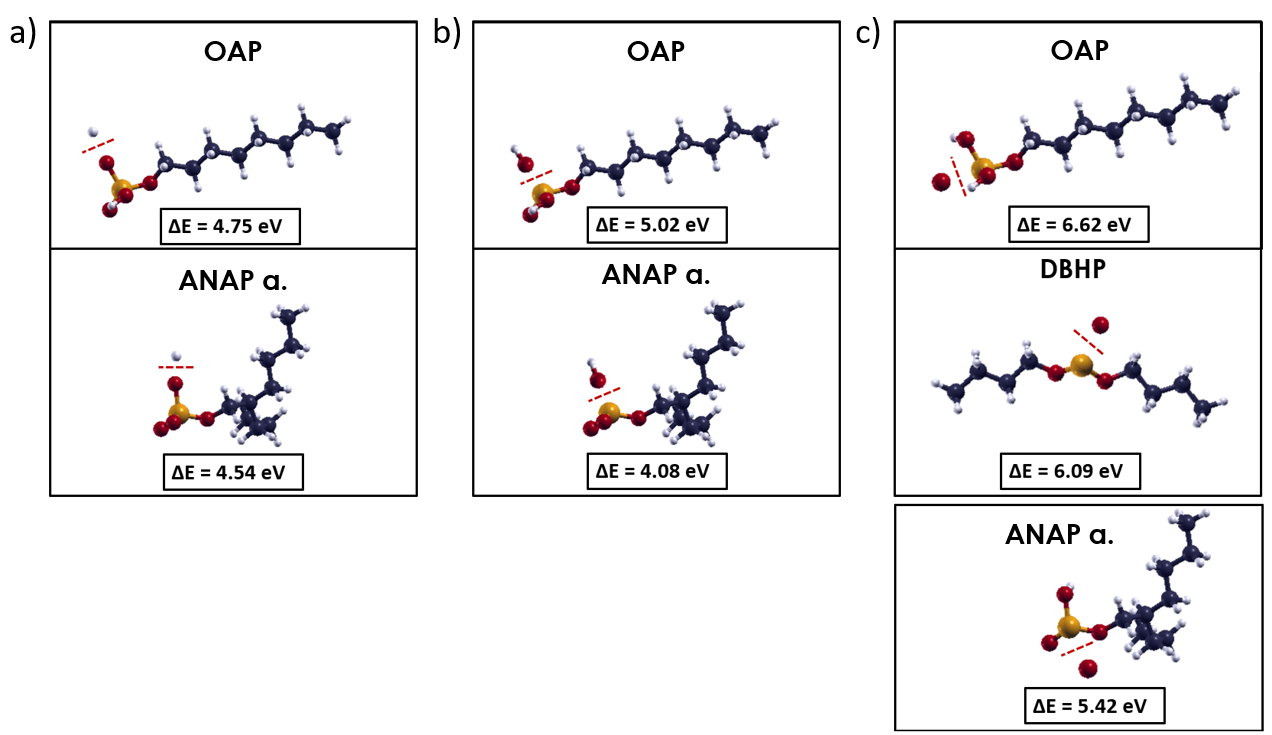}
\caption{\label{fig:stat} Energy costs associated to the removal of an H atom (a), an hydroxyl group (b) and an O atom (c) in the reported molecules.}
\end{figure}

To quantitatively confirm the different dissociation path between three- and four-fold coordinated phosphorous groups emerged in AIMD simulations, we evaluated the energy costs $\Delta E$, defined in Eq.~\ref{deltae}, to cut each atomic bond in the different molecules. 
These energy differences do not represent the energy barriers required for molecular dissociation, but rather an evaluation of the relative stability of different bonds. The cuts tested (alkyl, alkoxy, hydrogen, hydroxyl, and oxygen removal) and the $\Delta E$ values obtained are reported in Fig.~\ref{fig:chain} and Fig.~\ref{fig:stat}.

In agreement with the AIMD simulations, the OAP required the highest amount of energy to break its bonds for each of the cuts investigated. Fig.~\ref{fig:chain} shows two cuts corresponding to the alkyl- and alkoxy-chain removal. In both cases, the same cut requires the smallest amount of energy for DBHP (2.31 eV and 3.07 eV) and the highest for OAP (4.55 eV and 4.12 eV), suggesting a higher stability for the latter. Moreover, the alkoxy-chain removal across all molecules (Figs.~\ref{fig:chain}a-c) was less energetically costly than the alkyl removal (Figs.~\ref{fig:chain}d-e), with the exception for OAP. The removal of the alkoxy chains was always energetically less expensive than the \hbox{P--O} breaking shown in Fig.~\ref{fig:stat}c, suggesting that the head group subunits (like OH and H Figs.~\ref{fig:stat}a-b) were more stable and difficult to break compared to the alkoxy-chains. These findings could be relevant to understand which is the driving mechanism of tribofilm formation for other lubricant additives containing alkoxy chains~\cite{Philippon2007,De-Barros2015,Li2022} since this functional group seemed the easiest to break.

The higher bond strengths in OAP compared to other molecules was also confirmed by the bond order analyses reported in Tab.~\ref{tab:bondstrength}. The ICOBI values provided insight into the nature and strength of chemical bonds, identifying which ones are single or double-bonded. The different \hbox{P--O} bonds for the studied molecules are labelled as shown in Fig.~\ref{fig:mols}. In our case, we can indicate a single bond when ICOBI values range from 0.76 to 0.80, whereas values between 1.31 and 1.33 correspond to double bonds. As shown in the ICOBI columns of Tab.~\ref{tab:bondstrength}, the oxygen terminated bonds (\hbox{P--O1} for the OAP, \hbox{P--O2} for the DBHP) were always identified as double bonds, with the remaining ones classified as single bonds. In the phosphorus-based fragment of the ANAP, we classified two single and two double covalent bonds. The ICOBI values of the latter fall between those computed for double bonds in OAP and DBHP. (1.02 and 1.01 for P--O1 and P--O2 bonds, respectively). These intermediate values can be attributed to the charged nature of this fragment, which leads to a redistribution of the additional negative charge.

On the other hand, the ICOOP analysis allowed to evaluate the bond strengths of the P--O bonds across the different molecules, i.e., which bonds are stronger or weaker within the molecule. The ICOOP columns in Table~\ref{tab:bondstrength} shows that the strength for the \hbox{P--O} bonds in OAP, both single- and double-bonded, is consistently larger than in DBHP, indicating a stronger \hbox{P--O} interaction for the former. In contrast, the phosphate anion of ANAP exhibited a hybrid behaviour, with ICOOP values closer to DBPH for the double \hbox{P--O} bonds and a stronger interaction for single \hbox{P--O} bonds, similar to OAP. The ICOBI and ICOOP analyses supported the AIMD simulations, confirming that four-fold coordinated head groups had the strongest bonds, thus increasing their stability compared to three-fold coordinated groups and requiring the highest energy cost to dissociate.

\begin{table}[htpb]
\begin{center}
\caption{\label{tab:bondstrength} ICOBI and ICOOP values calculated for the different \hbox{P--O} bonds in DBHP, OAP and ANAP a. fragment. O atoms indexing refers to Fig.~\ref{fig:mols}.}
\begin{tabular}{ c c c c c c c }
\hline
\hline
\multirow{2}{*}{} & \multicolumn{2}{ c }{DBHP} & \multicolumn{2}{ c }{DOAP} & \multicolumn{2}{ c }{ANAP a.}\\
\hline
                   & ICOBI & ICOOP & ICOBI & ICOOP & ICOBI & ICOOP\\
\hline
  \hbox{P--O1} &  0.79   &  0.18   &  1.33   &  0.45   &  1.02   &  0.33\\
  \hbox{P--O2} &  1.31   &  0.34   &  0.77   &  0.23   &  1.01   &  0.33\\
  \hbox{P--O3} &  0.78   &  0.17   &  0.76   &  0.22   &  0.78   &  0.23\\
  \hbox{P--O4} &  --     &  --     &  0.79   &  0.25   &  0.80   &  0.25\\
\hline 
\end{tabular}
\end{center}
\end{table}

\subsection{Experimental correlation}
\begin{figure*}[htpb]
\includegraphics[width=\textwidth]{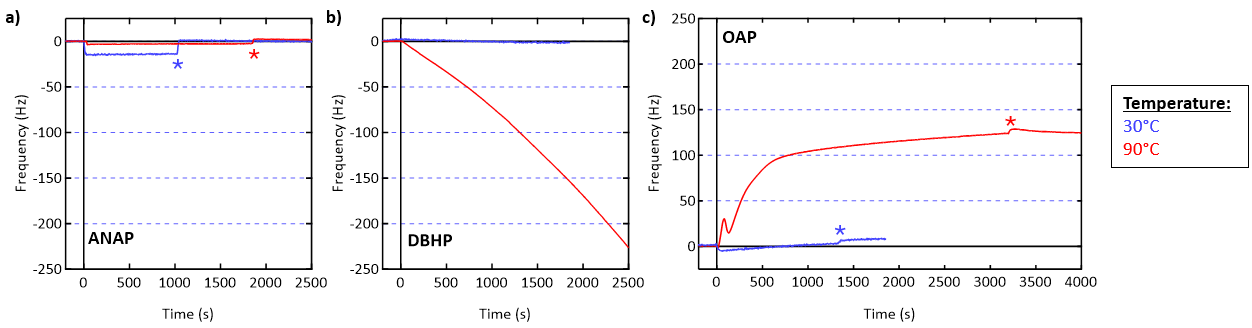}
\caption{\label{fig:qcm} QCM frequency variations for ANAP (panel a), DBHP (b) and OAP (c) at 30 °C (blue line) and 90 °C (red line). * symbols indicate the base oil time rinsing.}
\label{fig:QCM}
\end{figure*}

QCM measurements have been performed, for each solution, at temperatures of 30~°C and 90~°C respectively. Results are reported in Fig.~\ref{fig:qcm}. It must be noticed that the adsorption behavior is very different at these two temperatures and for the different phosphorus-based chemistry.

Concerning ANAP, shown in Fig.~\ref{fig:QCM}a, a small frequency drop is observed both at 30~°C (-14 Hz) and at 90~°C (-2.5 Hz). Besides, a complete desorption is observed when rinsed with base oil. These observations may result from a weak interaction between ANAP and the metallic surface due to a low physisorption interaction. Different frequency drops between 30~°C and 90~°C must come from a solubility difference of the molecule in the base oil. Indeed, at higher temperature, solubility will increase and will result in a decrease of interaction between the AW and the metallic surface.
For DBHP in Fig.~\ref{fig:QCM}b, a slow adsorption kinetics is observed at 30 °C, whereas at 90 °C a large and continuous frequency drop is measured. This behavior could result from the DBHP reactivity under temperature stress which leads to a large deposit formation. Note that for these two experiments, no base oil rinsing has been performed.
Finally, for OAP in Fig.~\ref{fig:QCM}c, a frequency increase is observed both at 30 °C and 90 °C. At higher temperature, the kinetic is faster than at ambient temperature. Usually, an increase in the QCM frequency shift is related to a loss of mass. After base oil rinsing step, frequency remains stable which means that the system is not going back to the initial state. To understand this mass loss, XPS analysis has been performed to characterize AW/substrate interface obtained at 90 °C (see Tab.~\ref{tab:exp}).

\begin{table}[htpb]
\begin{center}
\caption{\label{tab:exp} Elemental composition (at\%) obtained by XPS analysis (after base oil rinsing for ANAP and OAP deposit) on QCM substrate and adsorption performed at 90~°C.}
\begin{tabular}{c c c c c c c}
\hline
\hline
\textbf{System}                     & \textbf{C} & \textbf{N} & \textbf{O} & \textbf{Fe} & \textbf{P} & \textbf{P/Fe} \\ 
\hline
\textbf{Pristine stainless-steel} & 28.6       & 1.7        & 53.6       & 16.1        & -          & -             \\
\textbf{ANAP}                     & 31.6       & 1.2        & 53.7       & 10.2        & 3.3        & 0.32          \\
\textbf{DBHP}                     & 15.6       & 6.8        & 56.4       & 2.9         & 18.3       & 6.31          \\
\textbf{OAP}                      & 39.5       & 4.2        & 47.6       & 2.6         & 6.1        & 2.34          \\
\hline
\end{tabular}
\end{center}
\end{table}

First, it must be noted that an organic contamination layer is present on top of the raw substrate that is used for this study. Nitrogen atoms are also detected which may come from the stainless-steel deposit process. Moreover, the stainless steel reference showed a high concentration of oxygen, which suggests the presence of a iron oxide layer over the metal surface.
For ANAP deposit, elemental composition is close to the one of pristine stainless-steel surface. This result is in great agreement with the adsorption mechanism measured in QCM as a complete AW desorption is observed when rinsed with base oil. This result is also correlated with simulation data, where we identified a relevant molecular adsorption with no chemical reactivity under tribological conditions.
For DBHP deposit, a large amount of P is detected which confirms the high reactivity at 90~°C. Here again, the result is in great agreement with AIMD simulations as they have shown that phosphite exhibits the highest chemical reactivity.
For OAP deposit, as for DBHP, a P-based layer is observed which does not explain the mass loss observed on QCM data.
Moreover, the Fe amount detected via XPS is quite low with a P/Fe ratio above 1. This result shows that OAP additive adsorbs on the surface which must induce a mass increase that should be measured via QCM. Yet QCM data for OAP is not showing a mass adsorption, that is why further analysis on the substrate chemistry will be presented later. From XPS elemental analysis, it must be also noted that C/P ratio for OAP is equal to 6.4 (in theory 8 as presented in Figure 1) whereas for DBHP, the C/P ratio is equal to 0.9. This shows clearly that at high temperature, DBHP is degraded to form mainly a P-based layer at the surface, while losing the C chain.

\begin{figure*}[htpb]
\includegraphics[width=\textwidth]{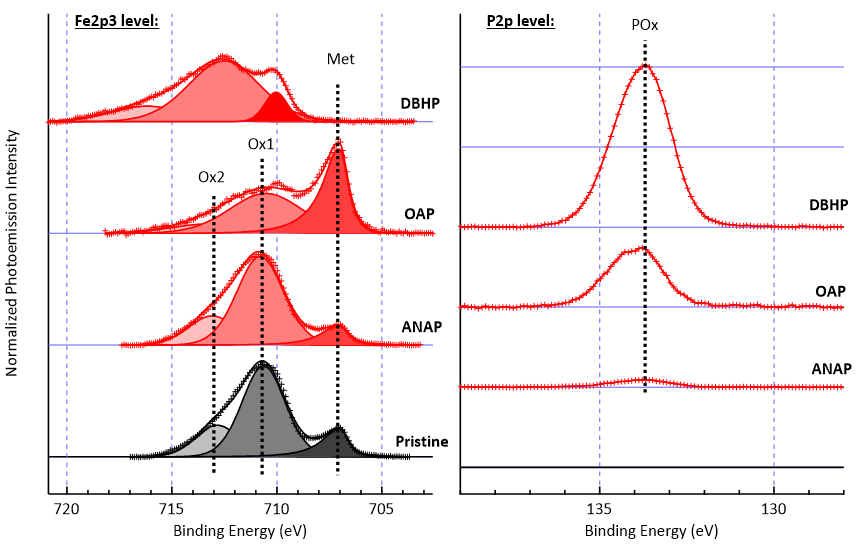}
\caption{\label{fig:xps} XPS measurements for the variation of the Fe2p3 and P2p levels on the pristine stainless steel reference (black and grey, lowest panel) and the three investigated molecules (red and pink).}
\end{figure*}

As simulation data for the OAP additive suggested the release of hydrogen atoms over the metal surface, an XPS analysis of the iron oxide layer variation has been performed, with the results shown in Fig.~\ref{fig:xps}.

The Fe2p3/2 peak measured via XPS clearly shows the different iron states within the stainless steel reference: 707 eV for metallic Fe and around 711 eV for oxidized Fe. We want to remark again, as shown in Tab.~\ref{tab:exp}, that even if the experimental work has been performed on a stainless-steel surface, the iron chemistry shows a higher amount of iron oxide, which is in agreement with a hematite-type top surface.
The interaction between the different AW additives and the substrate dramatically changed the shape of the Fe2p3/2 peaks: for example, in the stainless-steel and ANAP case, no differences were observed in the peak shape and position, which confirms that there is no significant reactivity for this additive over the substrate. It is worth noting that for DBHP, a shift is observed at Fe edge but not at P edge. This could be explained by a differential charge effect between the steel and the thick P-based layer surfaces. It also confirms the reactivity of DBHP under high temperature. This suggest a tendency to molecular dissociation promoted by temperature, a mechanism which might as well be enhanced by the harsh tribological conditions of load and shear simulated in the AIMD. Indeed, the molecule appeared to dissociate almost immediately in this simulation, leaving P chemisorbed on Fe as well as its two alkoxy chains which could explain both the high P content and the mass increase recorded in the QCM experiments.
Finally, for OAP, the peak associated with the oxidation state decreased while the corresponding metallic contribution increased compared to the reference surface. Thus, it indicates that this AW additive reacts with Fe oxide. This result is in agreement with the OAP adsorption that leads to a surface reduction and lose of oxide layer. The QCM behavior presented in figure Fig.\ref{fig:QCM}c is then the results of both mass loss from surface oxide and mass increase from the OAP adsorption: the mass loss being higher than the masse increase, the behavior globally conducts to a mass loss. Comparing this with the results obtained by \textit{ab initio} analysis, in which a release of H from the molecule occurred both in the adsorption on hematite and in the molecular dynamics simulation on iron, we can hypothesize a removal of the Fe oxide surface layer due to the reducing effect of H, explaining the mass loss associated to an increase in the frequency measured by QCM.

\section{Conclusions}

This study focuses on the adsorption and stability of three phosphorus-based lubricant additives, OAP, DBHP, and ANAP, on iron and hematite substrates. Through \textit{ab initio} calculations, we found that ANAP, followed by DBHP, exhibited the strongest interaction with bare iron, while OAP displayed a weaker adsorption on both substrates. The adsorption trends were consistent across Fe and hematite substrates, with a notable enhancement for OAP on hematite, where the interaction was facilitated by the loss of a hydrogen atom.

\textit{Ab initio} molecular dynamics simulations showed that DBHP dissociated more easily than OAP and ANAP when interacting with iron surfaces under tribological conditions. This effect was attributed to the lower phosphorus coordination of the phosphite molecule with respect to the phosphate ones, which promotes molecular dissociation. We confirmed these results with a bond order analyses, where we found the \hbox{P--O} bond in OAP has the highest strength, confirming that it was less prone to dissociation.

Experimental QCM and XPS analyses further validated our results. Notably, DBHP demonstrated the highest P/Fe ratio, consistent with its strong chemisorption and dissociative behavior observed in simulations. Conversely, OAP exhibited a loss in mass and reduction in Fe oxide peak, suggesting a removal of surface oxides through the release of hydrogen, as observed in the simulations.

Overall, this approach shows that \textit{ab initio} calculations can provide valuable information to understand the mechanisms of adsorption and reaction of additives on ferrous substrates. Indeed, the results here presented turned out to be in very good agreement with the experimental observations. In particular, we highlight how molecular structure and oxidation states of phosphorus additives affect molecular adsorption which is the first step for tribofilm formation, providing valuable insights for the design of lubricants tailored to prevent scuffing and enhance performance under diverse conditions.

\section*{Acknowledgements}
MCR, PR, and FB acknowledge the SLIDE project, which received funding from the European Research Council (ERC) under the European Union’s Horizon 2020 research and innovation program. (Grant Agreement No. 865633). They also acknowledge the CINECA award under the ISCRA initiative, for the availability of high-performance computing resources and support.

\section*{Data availability}
Data will be made available on request.

\bibliographystyle{elsarticle-num}
\bibliography{bibliography}% Produces the bibliography via BibTeX.

\end{document}